\documentclass[11pt,preprint,graphicx]{aastex}
\usepackage{graphicx}
\usepackage[labelfont={sf,bf}, margin=1cm]{caption}

\def\etal{\it et al. \rm }

\begin{document} 

\title{Systematic Bias in 2MASS Galaxy Photometry}

\author{James Schombert}
\affil{Department of Physics, University of Oregon, Eugene, OR 97403;
jschombe@uoregon.edu}

\begin{abstract}

\noindent We report the discovery of a serious bias in galaxy photometry reported in
the 2MASS Extended Source Catalog (Jarrett \etal 2000).  Due to an
undetermined flaw in the 2MASS surface photometry routines, isophotal and
total magnitudes calculated by their methods underestimate the luminosity
of galaxies from 10\% to 40\%.  This is found to be due to incorrectly
determined scalelengths and isophotal radii, which are used to define the
aperture sizes for Kron and total fluxes.  While 2MASS metric aperture
luminosities are correct (and, thus, colors based on those apertures),
comparison to other filters (e.g. optical) based on total magnitudes will
produce erroneous results.  We use our own galaxy photometry package
(ARCHANGEL) to determine correct total magnitudes and colors using the same
2MASS images, but with a more refined surface brightness reduction scheme.
Our resulting colors, and color-magnitude relation, are more in line
with model expectations and previous pointed observations.

\end{abstract}

\section{Introduction}

Surface photometry is an important tool in the understanding of galaxy
mass and structure.  Galaxy luminosity measures the primary baryonic
component, (i.e. stellar mass) and structural information traces the
gravitational potential.  Galaxy formation scenarios make specific
predictions on the light distribution of galaxies, so accurate reduction of
a galaxy image into a total magnitude and scalelength are important
parameters to understanding the fundamental plane and the star formation
history of galaxies.

Obtaining the structural characteristics and the total luminosity of a
galaxy requires knowledge of its surface brightness profile to a
significant depth.  In order to extrapolate a total luminosity, isophotal
analysis is required to determine how far one needs to integrate a galaxy's
light plus to provide sufficient information to extrapolate the light
profile.

During a surface photometry project to explore the structure of galaxies by
morphological type (Schombert \& Smith 2011), we discovered a significant
discrepancy between the structural and luminosity parameters that we
determined using raw 2MASS images versus those reported by the 2MASS project in
their Extended Source Catalog (Jarrett \etal 2000).  We use this letter to
outline the problem, and the solution, for other researchers.

\section{Sample}

Our original project was to perform surface photometry of large galaxies
over a full range of morphological types.  This type of analysis was
previously attempted by 2MASS (Jarrett \etal 2003), but their focus was not
on the structural characteristics of galaxies, but rather luminosities and
colors.  The first stage of our project was to understand the surface
photometry of ellipticals, the simplest galaxies for structural studies as
they have highly symmetric isophotes and very few complications to their
light distributions due to extinction or recent star formation.

\begin{figure}[!ht]
\centering
\includegraphics[scale=0.8,angle=0]{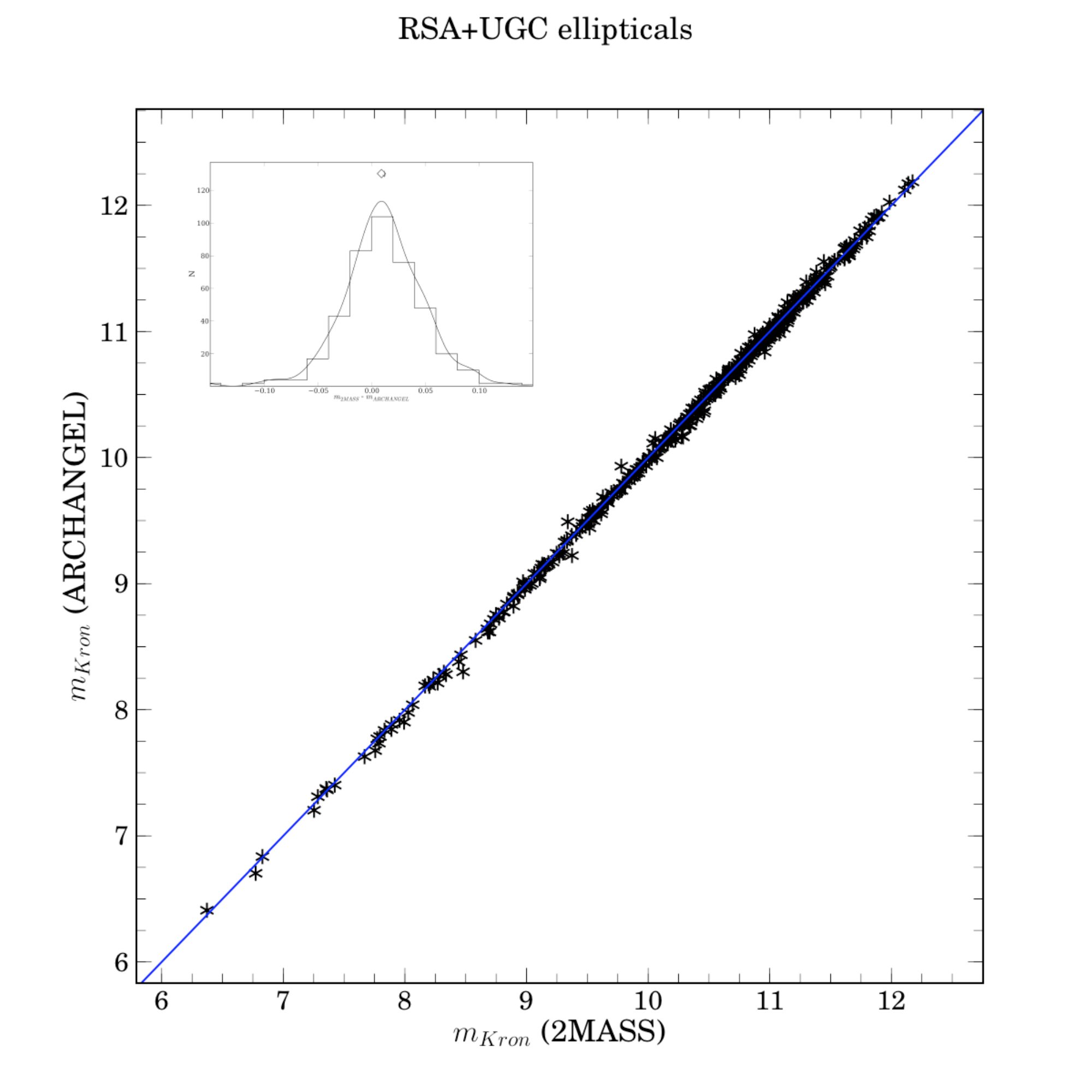}
\caption{\small Comparison of 2MASS $J$ Kron apparent magnitudes for 421 ellipticals
on our galaxy structure survey (Schombert \& Smith 2011).  The blue line is
the one-to-one equivalence line.  The agreement is excellent as we use the
aperture sizes and orientations given by the 2MASS project.  The inset
histogram is the difference in magnitudes, the mean difference is 0.01 mags
(our magnitudes are slightly fainter because of our pipeline reduction
procedures that subtract stars and replaces their pixels with interpolated
galaxy flux).
}
\end{figure}

Our sample was selected from the Revised Shapley-Ames (RSA) and Uppsala
Galaxy Catalogs (UGC) in order to cover a magnitude and angular limited
sample with sufficient S/N in the 2MASS image library.  Our only other
criteria was that the galaxies to be studied be free of nearby companions
or bright stars which might disturb the analysis of the isophotes to faint
luminosity levels.  Our final sample contained 421 galaxies all classed 'E'
by both catalogs.

We downloaded the 2MASS $J$, $H$, $K$ regions of the sky around all the
galaxies in the sample from 2MASS's Interactive Image Service.  These sky
images were flattened and cleaned by the 2MASS project and contained all the
information needed to produce calibrated photometry.  We analyzed the images
using our own galaxy photometry package (ARCHANGEL, Schombert 2007), thus,
the only difference in the final results is our analysis methods, not the
data themselves.

\section{2MASS Repeatability}

Our obvious first step, once we completed our surface photometry reduction,
is to compare our photometric and structural values with those extracted by
2MASS.  Metric magnitudes are the simplest for comparison.  The 2MASS
project provides magnitudes through various aperture sizes (e.g. 14 arcsecs
is found in NED) and we were able to reproduce their values in all three
bandpasses.

\begin{figure}[!ht]
\centering
\includegraphics[scale=0.8,angle=0]{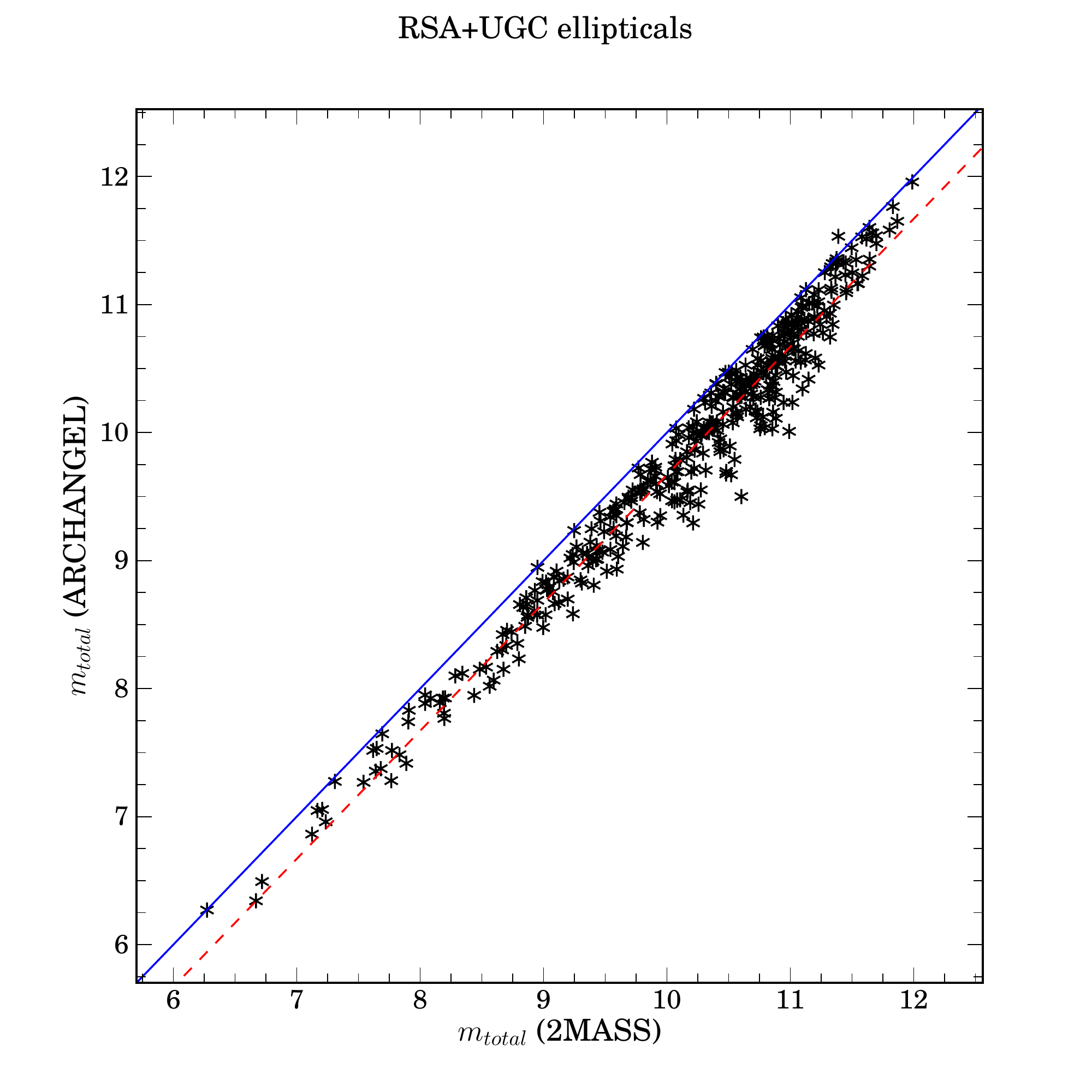}
\caption{\small Comparison to total magnitudes ($J$ band) from the 2MASS
Extended Source Catalog with our photometry from the 2MASS raw images.  The
blue line represents one-ton-one correspondence, the red line is a linear
fit with a slope of 1.001.  The 2MASS total
magnitudes are 0.33 mags fainter than our calculated total magnitudes.
This represents an error ranging from 10 to 40\% in total luminosity.
}
\end{figure}

The 2MASS project also provides Kron magnitudes.  Kron magnitudes are
isophotal magnitudes measuring a galaxy's light through an elliptical
aperture whose size is defined by the 20 $K$ mag arcsecs$^{-2}$ surface
brightness level.  These magnitudes contain less intrinsic error as the
Kron apertures follow the shape of the galaxy and maximizes the galaxy flux
to sky ratio.   NED provides those magnitudes and the aperture sizes for
all the galaxies in our sample.  We compare our Kron magnitudes (using
2MASS's aperture sizes) with their Kron magnitudes in Figure 1.

As can be seen from Figure 1, the agreement between our Kron magnitudes and
2MASS values is excellent, meaning that we can reproduce the same fluxes as
the 2MASS project using the same apertures.  There is a slight offset (0.01
mags) such that our magnitudes are slightly fainter than 2MASS (see inset
histogram).  This is probably due to the fact that we subtract stars and
replace the masked pixels with interpolated galaxy flux which, on average,
would lower the aperture flux.

\section{Problems with 2MASS Total Magnitudes}

Our next comparison was with our total magnitudes and 2MASS's total
magnitudes.  That comparison is found in Figure 2 and a significant
difference is found between our calculated total magnitudes and the values
presented for the same galaxies by the 2MASS project.  In general, our total
magnitudes are 10 to 40\% brighter than the 2MASS total luminosities.

\begin{figure}[!ht]
\centering
\includegraphics[bb=60 30 490 490,width=5.5in]{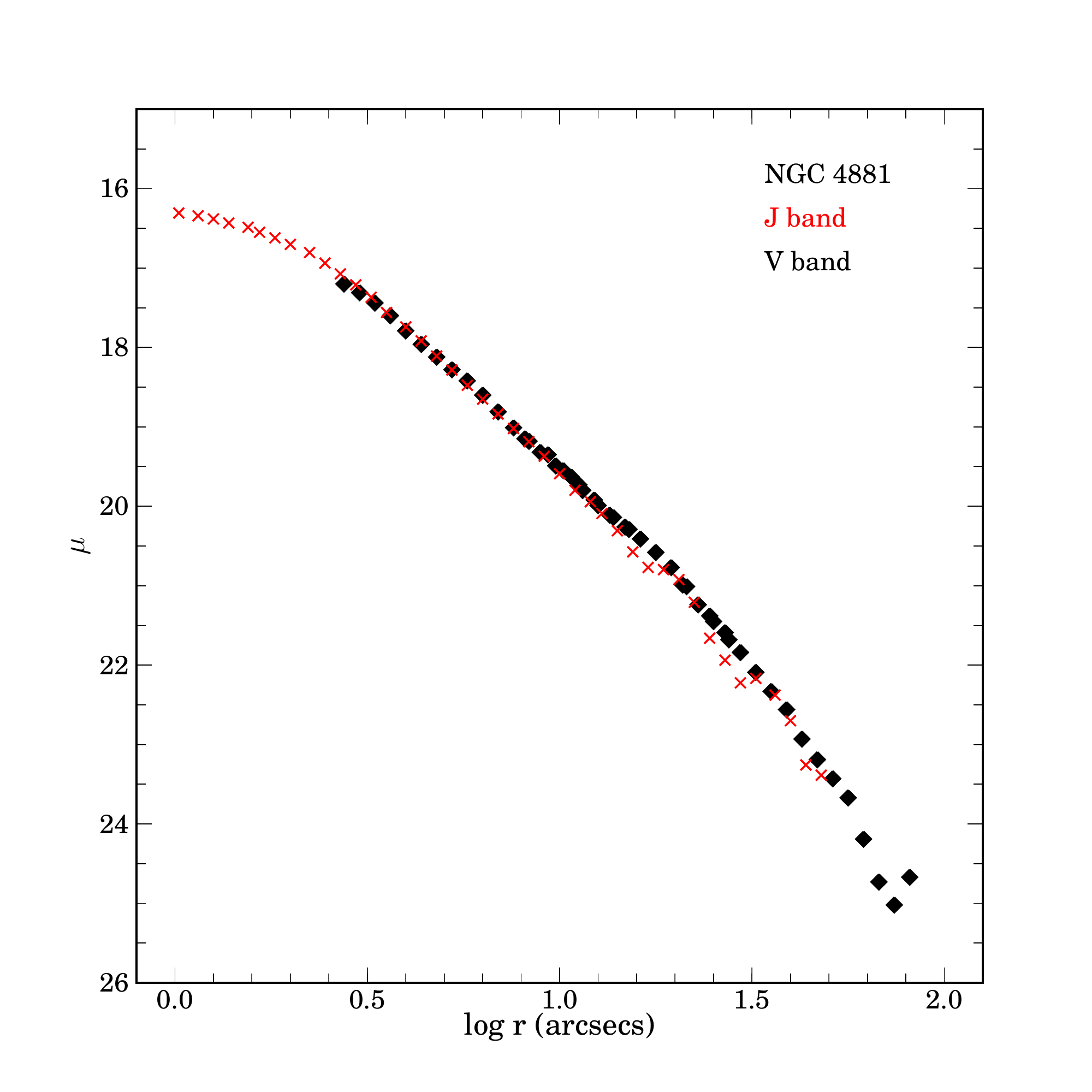}
\caption{\small A comparison of the $J$ surface brightness profile for NGC
4881 in the Coma cluster and old $V$ surface brightness data.  The $J$
profile is reduced from 2MASS images using our ARCHANGEL galaxy photometry
system.  The $V$ data is from photographic material used in Schombert
(1985) shifted for a $V-J$ color of 2.3.  The agreement is excellent
considering that the data is separated by 25 years, different wavelengths,
different detectors and different reduction software.}
\end{figure}

This discrepancy in total luminosities is especially puzzling since we can
reproduce 2MASS aperture and Kron magnitudes.  Thus, we believe the images
provided by the 2MASS project are reliable and the calibration is correct.
Total magnitudes determined by the 2MASS project use an aperture magnitude
that is four scalelengths in radius where the scalelength is determined
S\'{e}rsic fits to their surface brightness profiles.  Our project
determines total magnitudes through asymptotic fits to the curve of growth,
where we increase the S/N of the outer isophotes by using the mean
intensities give by the surface brightness profiles.

The key difference in our photometric methodology lies in the determination
and use of each surface brightness profile.  We suspect the difference is
because of the methodology used by the 2MASS project, compared to our
technique.  To explore this hypothesis, we compare the procedures used by
2MASS and ourselves in the next sections in the hope of locating the difference.

\subsection{Surface Photometry Comparison}

Our procedure for isophotal analysis of galaxies is found in Schombert
(2007), and the full data for the galaxies used in this letter will be
presented in Schombert \& Smith (2011).  Our procedure is common to galaxy
surface photometry.  Ellipses are fit to the isophotes then surface brightness
profiles are then constructed from the mean intensities around those
ellipses.  An example of our final results is shown in Figure 3, a $J$
profile from 2MASS images and $V$ profile from old photographic information
(Schombert 1985, corrected for a $V-J$ color of 2.3).

\begin{figure}[!ht]
\centering
\includegraphics[bb=70 220 490 590,width=5.5in]{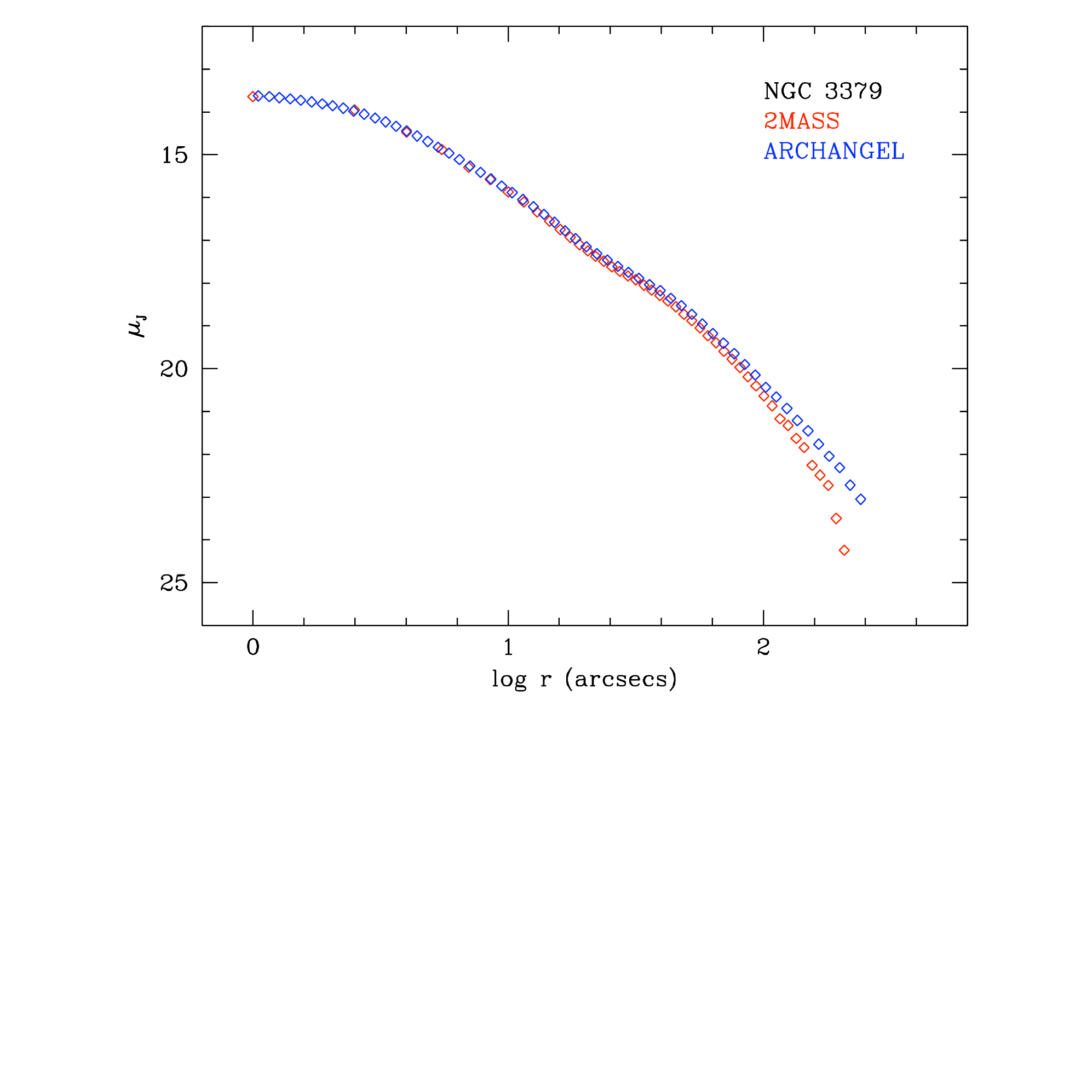}
\caption{\small A comparison of the $J$ surface brightness profile
presented by the 2MASS project (Jarrett \etal 2003) and the profile reduced by
our software package (ARCHANGEL).  The photometry agrees at high surface
brightnesses, but begins to disagree below 18 $J$ mag arcsecs$^{-2}$.  As
discussed in the text, the difference can not be explained by poor ellipse
fitting, calibration error or an improper sky value.
}
\end{figure}

The 2MASS project also published surface brightness profiles for 100 large
galaxies (Jarrett \etal 2003), 31 of them in common with our elliptical
sample.  Agreement between our surface brightness profiles and the 2MASS
project's profiles is less than adequate.  A comparative example is found
in Figure 4, the surface brightness profiles of NGC 3379 from Jarrett \etal
and our study.  The difference between the profiles is extreme at large
radii, well beyond expectations from the RMS errors in the data.

\begin{figure}[!ht]
\centering
\includegraphics[bb=60 30 490 500,width=5.5in]{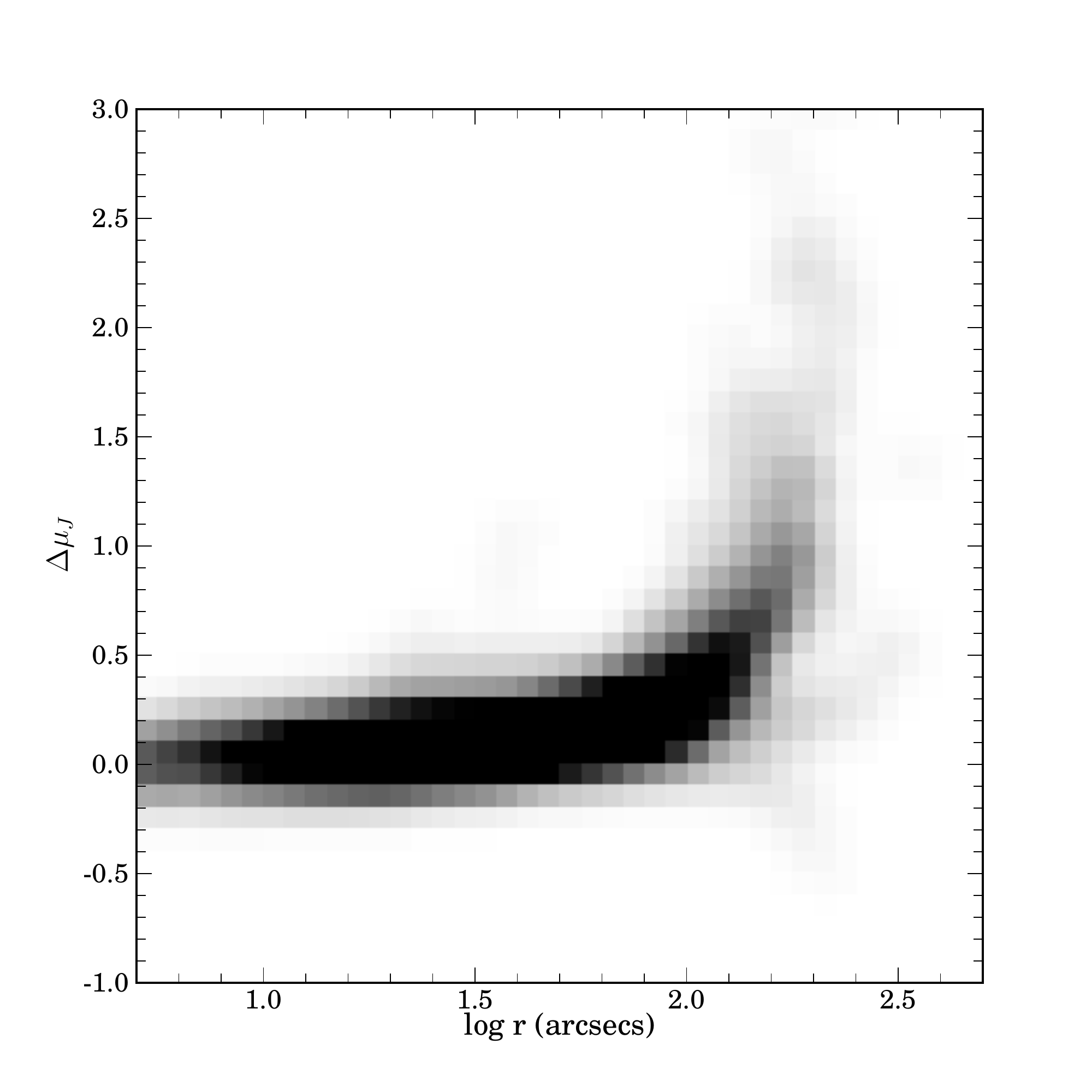}
\caption{\small A density plot of the surface brightness profile differences
between the 2MASS project and our study for 421 elliptical galaxies.  The
differences are primarily found in the outer regions, increasing with
galaxy radius.  The differences are uncorrelated with the luminosity of the
galaxy, size or any other physical characteristic that we can determine.
}
\end{figure}

The discrepancy for NGC 3379 is not unique.  The profile differences for
all 31 galaxies is shown in Figure 5, presented as a density distribution
of $\Delta\mu$ versus radius.  As can be seen in that Figure, all the
comparison galaxies have varying degrees of surface brightness differences,
mostly concentrated in the outer regions and can reach 1 to 2 mags
arcsecs$^{-2}$ in error.

\subsection{Data Reduction Differences}

One obvious conclusion is that some difference exists in the reduction
process that reflects in the final profiles, the data frames themselves are
not in question since we can reproduce 2MASS's aperture luminosities.
There are several procedural differences between the isophotal techniques
used by the 2MASS project and our photometry package (ARCHANGEL).

\begin{figure}[!ht]
\centering
\includegraphics[bb=70 220 490 590,width=5.5in]{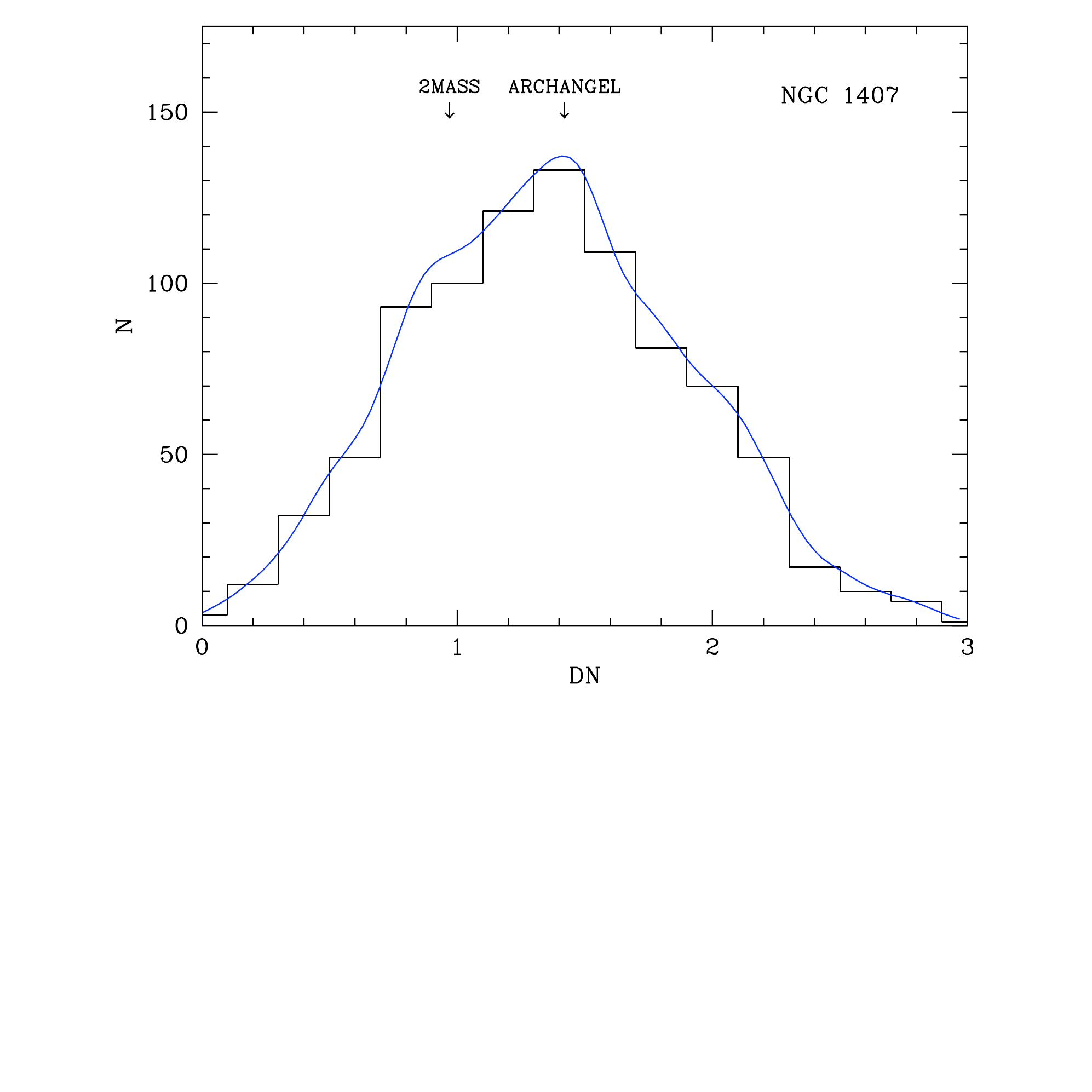}
\caption{\small A histogram of intensity values (in raw data units) for an annulus of
100 arcsecs (width of one pixel) for NGC 1407.  The 2MASS project cites a
value of 0.97 for this annulus, our study finds an intensity of 1.42.  The data
clearly supports our higher value.  This type of test was completed for all
31 galaxies in our surface brightness overlap sample, all produced the same
result.
}
\end{figure}

First, the 2MASS project determines an elliptical shape based on a first
moment analysis of some intermediate, but high S/N region in a galaxy's
envelope.  The calculated eccentricity and position angle are used for the
entire galaxy, determining isophote intensity levels based on pixels around
those ellipses.  Our project, on the other hand, fits each radii for
eccentricity and position angle (as well as x and y center) allowing these
ellipse parameters to vary with radius.

This difference in ellipse shape was noted in Schombert (2007), but these
difference ellipses are not enough explain the large surface brightness
differences found in the galaxy sample (numerical experiments with ellipses
in 2MASS data displays only a 1 to 2\% difference in intensities).  There
are a few extreme cases (e.g. LSB galaxy, NGC 3109), but in general
ellipticals have fairly constant eccentricities.  However, there are large
differences in the quoted intensity values per radius between the 2MASS
project and our study.  These differences range from small to up to 60\%,
greatest at the lowest intensity values.

\subsection{NGC 1407: Test Particle}

To resolve the differences in the surface brightness profiles, we selected
the elliptical NGC 1407 for detailed inspection.  NGC 1407 is an excellent
test particle for its isophotes are nearly circular (axial ratio of 0.93
from 2MASS, 0.95 from our study) and its envelope is free of any foreground
stars or distortions.

\begin{figure}[!ht]
\centering
\includegraphics[bb=70 220 490 590,width=5.5in]{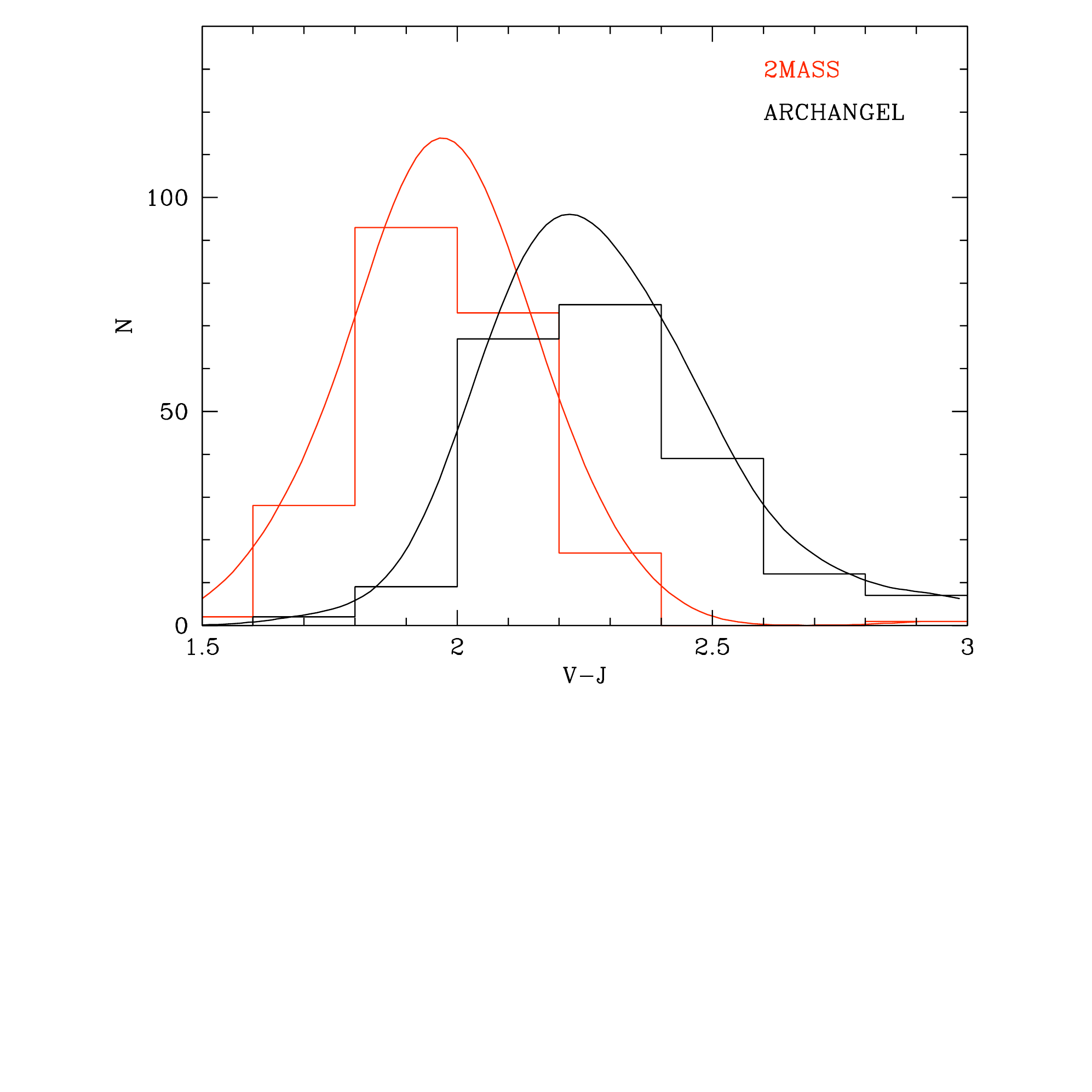}
\caption{\small A histogram of $V-J$ colors using galactic extinction
corrected total magnitudes from 2MASS (extracted from NED) and our study.
Since the 2MASS project underestimates the total magnitudes, this
reflects into bluer $V_J$ compared to our colors.  SED models predict a
$V-J$ color of 2.5 for a solar metallicity stellar population with an age
of 13 Gyrs, in-line with our colors.
}
\end{figure}

At 100 arcsecs from the center of NGC 1407, the 2MASS project quotes an
isophotal intensity of 0.97 DN (20.82 $J$ mag arcsecs$^{-2}$).  Our project
finds a value of 1.42 DN (20.41 $J$ mag arcsecs$^{-2}$).  To determine
which value more closely represents the isophote at that radius, we have
plotted a histogram of intensity values for all pixels between 99.5 and
100.5 arcsecs from the galaxy center.  This histogram is shown in Figure 6
(both regular and normalized).

From this Figure, it is obvious that the intensity values deduced by the
2MASS project are not in agreement with the mean value of the pixels in the
image, whereas our calculated intensity value is in good agreement with the
mean and median value.  Since NGC 1407 is a nearly perfect circle in axial
ratio, this is not an effect of the ellipse fitting procedure.  This is
also not due to calibration errors (these are raw data numbers) nor an
improper sky subtraction (the differences would be constant with radius,
they are not).  We are at a loss to explain 2MASS's values, however, our
values agree with the images (Figure 6) and with past photometry (Figure
3).

\begin{figure}[!ht]
\centering
\includegraphics[bb=70 220 490 590,width=5.5in]{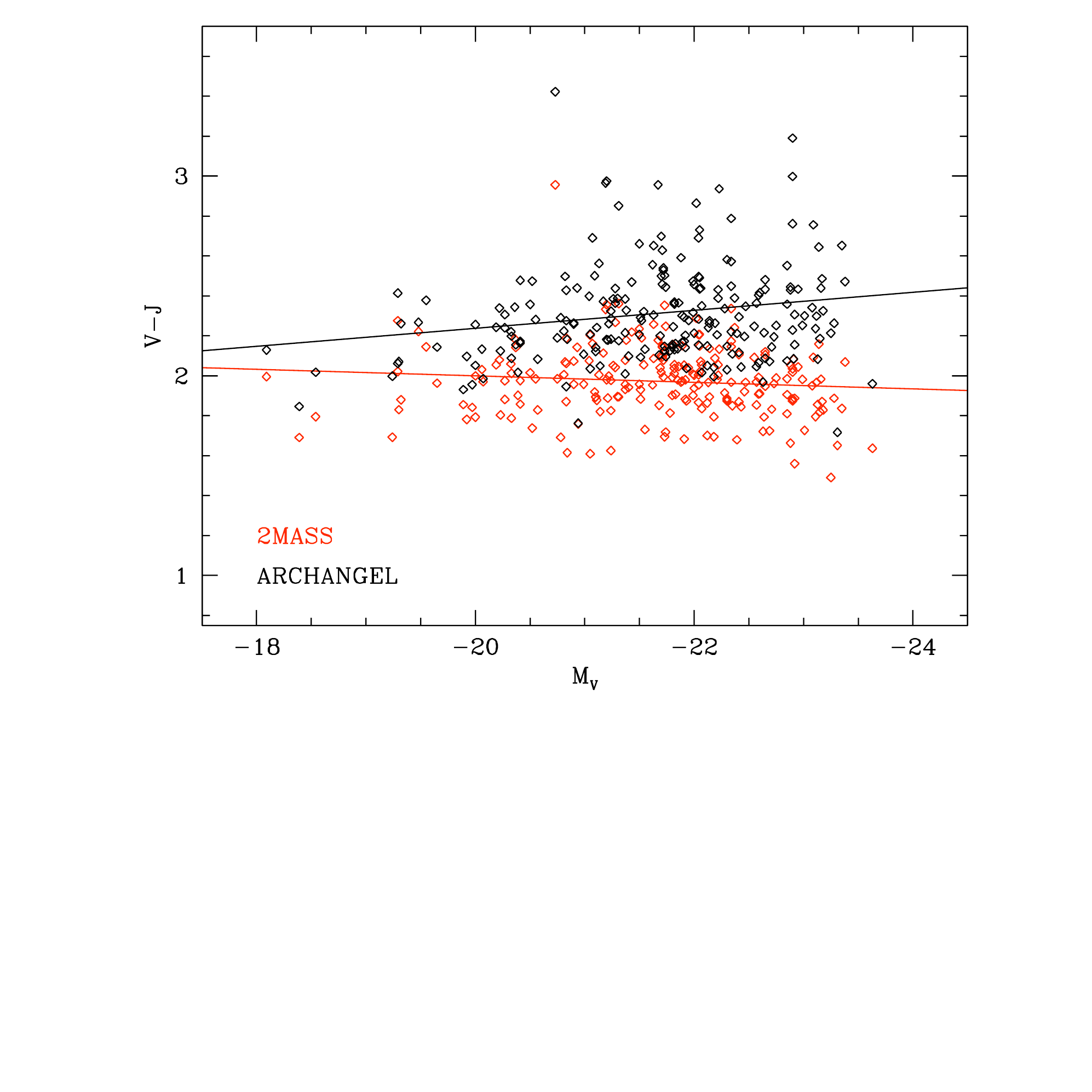}
\caption{\small The $V-J$ color-magnitude diagram for 2MASS colors (red
symbols) versus our study (black symbols).  Linear fits are shown.  The
2MASS project predicts a positive CMR slope, in contradiction with known
negative slopes in the literature.  Our study finds a negative slope
(redder colors with higher galaxy mass, i.e., higher mean metallicity).
}
\end{figure}

The effect these underestimated surface brightness values have on 2MASS
photometry is subtle.  Both 2MASS Kron and total magnitudes use the surface
brightness profiles to deduce isophotal levels (Kron) and scalelengths
(total).  For Kron magnitudes, the 20 $K$ mag arcsecs$^{-2}$ level is used
to define an elliptical aperture.  However, since the 2MASS surface brightness
profiles underestimate the intensity values per radius, this, in turn,
leads to smaller estimates of the isophotal size of the aperture and,
therefore, fainter magnitudes.  

Total magnitudes for 2MASS are calculated using an outer aperture set to be
four times the scalelength determined by S\'{e}rsic function fits.
Decreased intensities in the outer regions produce smaller scalelengths, on
average, which produce smaller apertures and fainter total magnitudes.
This is exactly what we observe in Figure 2.

\section{Summary}

First, we note that this discrepancy has no impact on projects which use
2MASS aperture colors.  For galaxy colors are calculated using 2MASS total
magnitudes still use the same sized apertures for $J$., $H$ and $K$, and
the colors will remain consistent (although for a smaller portion of the
total galaxy light).  However, comparison between other total magnitudes
(e.g. RC3 magnitudes) and 2MASS Kron or total magnitudes will be biased
towards the blue.

An example of this is shown in Figure 7, a histogram of $V-J$ colors for
the 421 ellipticals in our sample.  As can be seen, the 2MASS colors are
0.25 mags bluer than colors calculated from our total magnitudes since the
RC3 $V$ magnitude is determined from an asymptotic fit and, therefore,
contains more flux that 2MASS's total magnitude.   There appears to be no
standard correction from 2MASS colors to the correct colors, this would
require information on how deviant the 2MASS surface brightness profiles
(from which the aperture sizes are extracted) are from reality.  There
appears to be no simple formula.

A priority science goal for 2MASS was large baseline color comparison.  An
example of relevance of large wavelength comparisons is the color-magnitude
relation (CMR).  The CMR is a long known correlation between galaxy color
and luminosity.  The best explanation is that galaxies with higher mass
have higher metallicities.  Global metallicity reflects in the mean
temperature of the RGB such that low metallicities produce bluer colors.
The CMR for this data sample is shown in Figure 8.  Again, we see that the
2MASS colors predict the {\it opposite} expectation from earlier optical
work in that they find roughly bluer colors with higher luminosity.  Using
our total magnitudes (combined with RC3 colors) restores the correct CMR,
redder colors with higher galaxy luminosity.

We summarize our findings as the following:

\begin{description}

\item{(1)} 2MASS aperture magnitudes are reproducible with high accuracy
when the same metric apertures are used regardless of shape.

\item{(2)} Comparison between our photometry packages total or Kron
magnitudes and 2MASS's values (from the Extended Source Catalog, Jarrett
\etal 2000) find serious bias in the 2MASS luminosities of between 10 to
40\%.

\item{(3)} Point by point comparison of the surface brightness profiles
provided by Jarrett \etal (2003) indicate that problem lies in the
intensity values deduced by each photometry pipeline.  2MASS values are
increasingly fainter with increasing radius, this results in surface
brightness profiles that are smaller in isophotal radius compared to our
profiles.

\item{(4)} Comparison to raw intensity values (e.g., the 100 arcsecs
annulus in NGC 1407) demonstrates that, for unknown reasons, the 2MASS
project extracts lower intensity values than given by the images.  This is
either a flaw in the processing pipeline or an error in the data storage
system.

\item{(5)} Underestimated intensity values produces smaller isophotal
radii at constant surface brightness and smaller scalelength fits.  Both
result in smaller apertures and underestimated luminosities.

\item{(6)} While logically, either our photometry or 2MASS's is in error,
we consider three pieces of evidence to indicate our values are correct.
First, the direct comparison of intensity values from the raw images.
Second, the final $V-J$ colors where our colors are in agreement with mean
elliptical colors in the literature, the 2MASS project's values are 0.25
too blue.  Third, there is a negative slope in the CMR from 2MASS data,
ours data correct matches the color slope expected from SED models and
optical CMR's.

\end{description}

It is highly confusing on why this discrepancy has not been noticed to
date.  Either we have become too trusting of electronically available datasets.
Or we simply, due to pressures to publish, are not as thorough in our data
analysis, especially with respect to external checks.  We becoming
increasingly vulnerable to systematic errors with increasing automation to
our datasets, leading to potentially highly embarrassing errors in our
science results.

\noindent Acknowledgements: Most of the comparative values were extracted from NED
(NASA's Extragalactic Database) using new network tools.  The quick access
to difference galaxy catalogs on one site made this project doable in
reasonable timescales.  The model for future science is not faster cycles,
but faster and clearer access.  The software for this project was supported
by NASA's AISR and ADP programs.

\end{document}